\documentclass[a4paper, 11pt]{article}

\usepackage[swedish,english]{babel}
\usepackage{enumerate}  
\usepackage{caption}   
\usepackage{lscape}
\usepackage{amsmath,mathrsfs}
\usepackage{amssymb}  
\usepackage{epstopdf}
\usepackage{cite}
\usepackage{natbib}
\usepackage{subcaption}
\usepackage[colorlinks=true, linkcolor=black, citecolor=black, urlcolor=black]{hyperref}
\usepackage{nicefrac}
\usepackage{graphicx}
\usepackage{bm}	
\usepackage{algorithm}
\usepackage{algpseudocode}
\usepackage{algpascal}
\usepackage{booktabs}
\usepackage{multicol}
\usepackage[x11names]{xcolor}
\usepackage{lipsum}
\usepackage{geometry}
\usepackage{tikz}
\usetikzlibrary{tikzmark,calc,decorations.pathreplacing}

\newcommand{\ie}{i.e.,\ }
\newcommand{\eg}{e.g.,\ }

\newcommand{\wrt}{w.r.t.\ }

\newcommand{\ceil}[1]{\lceil {#1} \rceil}

\newcommand{\model}[1]{{\fontfamily{qpl}\selectfont {\small #1}}}
\newcommand{\dataset}[1]{{\fontfamily{ptm}\selectfont {\small #1}}}

\newcommand{\paperDTitle}{MAMOC: MRI Motion Correction via Masked Autoencoding}
\title{\paperDTitle}
\date{}

\begin{document}

\maketitle
\begin{center}
  Lennart Alexander Van der Goten$^{1,2}$ \\
  \texttt{lavdg@kth.se} \\[1ex] 
  Jingyu Guo$^{1,2}$ \\
  \texttt{jingyug@kth.se} \\[1ex]
  Kevin Smith$^{1,2}$ \\
  \texttt{ksmith@kth.se} \\[2ex]
  \emph{for the Alzheimer's Disease Neuroimaging Initiative}$^\ddagger$
\end{center}

\begin{center}
    \textsuperscript{1}KTH Royal Institute of Technology, Stockholm, SWEDEN\\
    \textsuperscript{2}Science for Life Laboratory, Solna, SWEDEN
\end{center}

\begin{abstract}
The presence of motion artifacts in magnetic resonance imaging (MRI) scans poses a significant challenge, where even minor patient movements can lead to artifacts that may compromise the scan's utility.
This paper introduces \textbf{MA}sked \textbf{MO}tion \textbf{C}orrection (\model{MAMOC}), a novel method designed to address the issue of Retrospective Artifact Correction (RAC) in motion-affected MRI brain scans. 
\model{MAMOC} uses masked autoencoding self-supervision, transfer learning and test-time prediction to efficiently remove motion artifacts, producing high-fidelity, native-resolution scans.
Until recently, realistic, openly available paired artifact presentations for training and evaluating retrospective motion correction methods did not exist, making it necessary to simulate motion artifacts.
Leveraging the \dataset{MR-ART} dataset and bigger unlabeled datasets (\dataset{ADNI}, \dataset{OASIS-3}, \dataset{IXI}), this work is the first to evaluate motion correction in MRI scans using real motion data on a public dataset, showing that \model{MAMOC} achieves improved performance over existing motion correction methods.
\end{abstract}

\begin{figure}[t]
  \centering
  \begin{minipage}{.24\linewidth}
      \centering
    (a)\\
    \includegraphics[width=\linewidth]{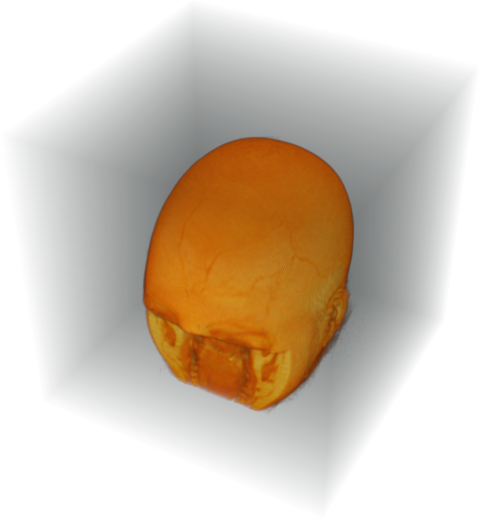}
  \end{minipage}\hfill
  \begin{minipage}{.24\linewidth}
      \centering
    (b)\\
    \includegraphics[width=\linewidth]{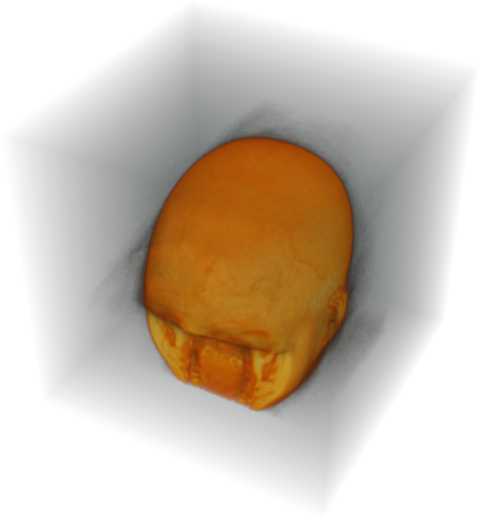}
  \end{minipage}\hfill
  \begin{minipage}{.24\linewidth}
      \centering
    (c)\\
    \includegraphics[width=\linewidth]{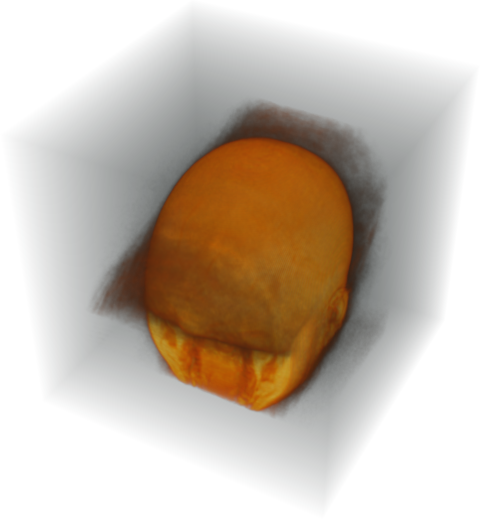}
  \end{minipage}\hfill
  \begin{minipage}{.24\linewidth}
      \centering
    (d)\\
    \includegraphics[width=\linewidth]{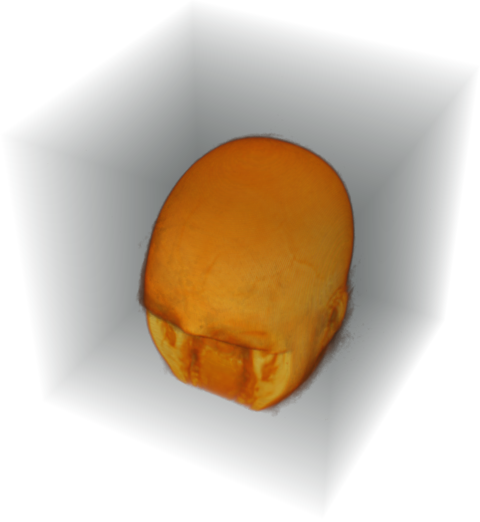}
  \end{minipage}
  \bigskip 
  \begin{minipage}{.24\linewidth}
    \includegraphics[width=\linewidth]{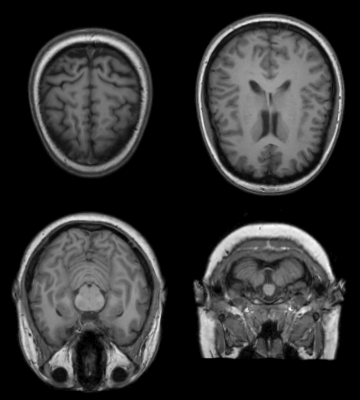}
  \end{minipage}\hfill
  \begin{minipage}{.24\linewidth}
    \includegraphics[width=\linewidth]{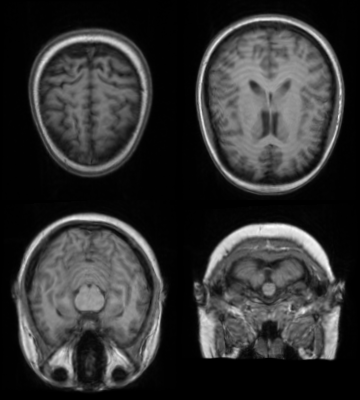}
  \end{minipage}\hfill
  \begin{minipage}{.24\linewidth}
    \includegraphics[width=\linewidth]{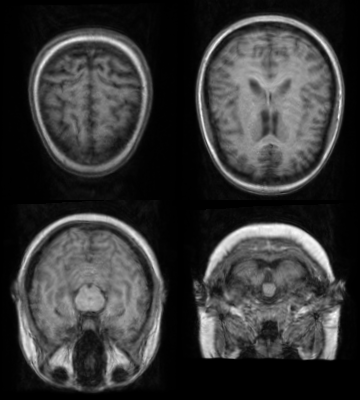}
  \end{minipage}\hfill
  \begin{minipage}{.24\linewidth}
    \includegraphics[width=\linewidth]{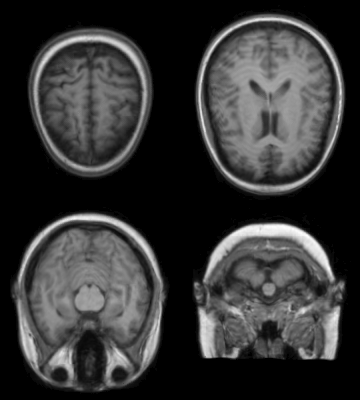}
  \end{minipage}
  \caption{\textbf{Varying Levels of Head Motion.} 3D \& 2D renderings of the same subject from the \dataset{MR-ART} dataset  with (a) no head motion, (b) moderate head motion and (c) heavy head motion, (d) motion-corrected with \model{MAMOC} }
  \label{PAPER-D:fig:motionCorrectionOverview}
\end{figure}

\section{Introduction}
\renewcommand{\thefootnote}{\fnsymbol{footnote}}
\footnotetext{$^\ddagger$ Data used in preparation of this article were obtained from the Alzheimer’s Disease
Neuroimaging Initiative (ADNI) database (\url{adni.loni.usc.edu}). As such, the investigators
within the ADNI contributed to the design and implementation of ADNI and/or provided data
but did not participate in analysis or writing of this report. A complete listing of ADNI
investigators can be found at:
\url{http://adni.loni.usc.edu/wp-content/uploads/how_to_apply/ADNI_Acknowledgement_List.pdf}}
\renewcommand{\thefootnote}{\arabic{footnote}}

Artifacts pose a serious challenge when it comes to magnetic resonance imaging (MRI). 
Performing an MRI scan is usually a costly endeavor as it requires specially-trained personnel, expensive highly-advanced medical equipment, and long scan durations. 
Commonly occurring artifacts (\eg motion, ghosting, bias fields, etc.) can cause a scan to become impossible to read, forcing healthcare providers to perform the procedure anew. 
Being able to filter out or correct artifacts (see Fig. \ref{PAPER-D:fig:motionCorrectionOverview}) is therefore of particular interest as it not only saves costs but also reduces discomfort in patients, especially when the artifact cannot be prevented (\eg due to Parkinson's disease, stroke, cardiac motion).

In this work, we propose  \model{MAMOC} (\textbf{MA}sked \textbf{MO}tion \textbf{C}orrection), which tackles the Retrospective Artifact Correction (RAC) task specifically on \emph{motion-affected} 
MRI scans.
It leverages the novel \dataset{MR-ART}  \citep{narai_movement-related_2022}  corpus, the first-of-its-kind dataset to provide multiple MRI brain scans per patient 
taken during the same session, with moderate motion, heavy motion, and a clean scan.
The small size of \dataset{MR-ART}, $148$ patients, makes it challenging to directly train a supervised model to correct motion artifacts, but it still provides unique value as a tool to evaluate RAC methods on realistic motion data instead of simulations.
To overcome the size limitation, we use self-supervision driven by \emph{masked autoencoding}  \citep{he_masked_2021} to learn the manifold of motion-unaffected data from several large, highly-curated datasets: \dataset{ADNI}  \citep{Wyman2013}, \dataset{OASIS-3}  \citep{LaMontagne} and \dataset{IXI}  \citep{ixi_dataset_2023}.
We then apply transfer learning on the limited labeled information available from the available from \dataset{MR-ART}.

Our contributions can be listed as follows:
\begin{enumerate}
    \item We propose a novel approach to remove motion artifacts from MR imagery using masked autoencoding self-supervision and transfer learning, \model{MAMOC}.
    \item We show that masked autoencoding in \model{MAMOC} can be used at inference time to improve the quality of reconstruction.
    \item \model{MAMOC} is able to synthesize native resolution motion-corrected MR imagery corresponding to voxel sizes of $1 \text{mm}^3$, $256^3$ voxels.
    \item We compare \model{MAMOC} to existing methods on \emph{real motion artifacts} from \dataset{MR-ART}, and demonstrate state-of-the-art performance \wrt image quality metrics, structural similarity, and a downstream subcortical segmentation task.
\end{enumerate}

\section{Related Work}
\subsubsection*{2D Image Restoration \& Generative Modeling}
Image restoration is a type of inverse problem  concerned with the automatic removal of artifacts from imagery  \citep{Zhang2017-gp, DBLP:journals/corr/abs-1803-04189}. 
Many modern works in this domain follow a holistic approach and are able to remove a wide class of artifacts  \citep{liang2021swinir, DBLP:journals/corr/abs-1803-04189} from imagery. 
While early research relied on handcrafted features, more recent approaches often adopt fully-learnable generative models  \citep{wu2021motion, https://doi.org/10.1002/mrm.27783}. 
In some recent works  \citep{liang2021swinir, 10.5555/3618408.3619991}  established convolutional elements have been supplanted with transformer-based architectures, allowing them to model larger contexts which can be advantageous for removing certain types of artifacts.

\subsubsection*{Prospective Artifact Correction}
Prospective Artifact Correction (PAC) refers to approaches that are concerned with either preventing artifacts altogether or by gathering sufficient information to remove artifacts in real-time  \citep{https://doi.org/10.1002/mrm.24314}.
Motion artifacts can be minimized by encouraging patients to remain still during the scan through clear instructions and by providing a comfortable scanning environment. 
In some cases, light sedation or other medical interventions  \citep{spieker_deep_2023} may be necessary, but these should be used judiciously and with the patient's well-being in mind.
Sensoric methods  \citep{zaitsev2006magnetic, herbst2011practical} utilize additional data sources to dynamically correct motion artifacts, offering a non-invasive alternative to physical restraints.
They typically assume motion rigidity and may extend scan durations, potentially increasing patient discomfort  \citep{https://doi.org/10.1002/mrm.24314}.

\subsubsection*{Retrospective Artifact Correction}
Retrospective Artifact Correction (RAC) methods seek to remove artifacts \emph{after} acquisition without relying on additional data  \citep{spieker_deep_2023}. 
RAC methods can typically be divided into two subfields: (i) $k$-space based and (ii) image-based. 
The former category operates directly in the $k$-space -- the frequency space which is the precursor to MR image space. 
The latter category takes motion-affected scans (post-reconstruction) as input and aims to produce a scan that features fewer pronounced artifacts. 
Given the limited availability of $k$-space data  \citep{noordman2023complexities}, our approach focuses on the image-based domain. 
Image-based works can be further categorized based on the generative modeling approach they utilize, such as Variational Autoencoders (VAEs)  \citep{https://doi.org/10.1002/mrm.27783} and GANs  \citep{wu2021motion}. 
In addition to generative models, supervised methods  \citep{9405626, vandergoten2022wide} utilize motion artifact simulators to generate paired scans to restore the image. 
These methods leverage the availability of simulated data to train models for specific tasks, thereby providing a practical solution to the challenges posed by the absence of paired real-world data.
\model{MAMOC} relies on a very limited amount of supervised training data containing \emph{real} motion artifacts, not simulations. 
Because the amount of supervised data is too small to train a model from scratch, \model{MAMOC} uses a self-supervised pre-training strategy that leverages large-scale, highly-curated datasets, to learn features for generating high-quality MRI imagery.

\begin{figure}[t]
\centering
\includegraphics[width=\textwidth]{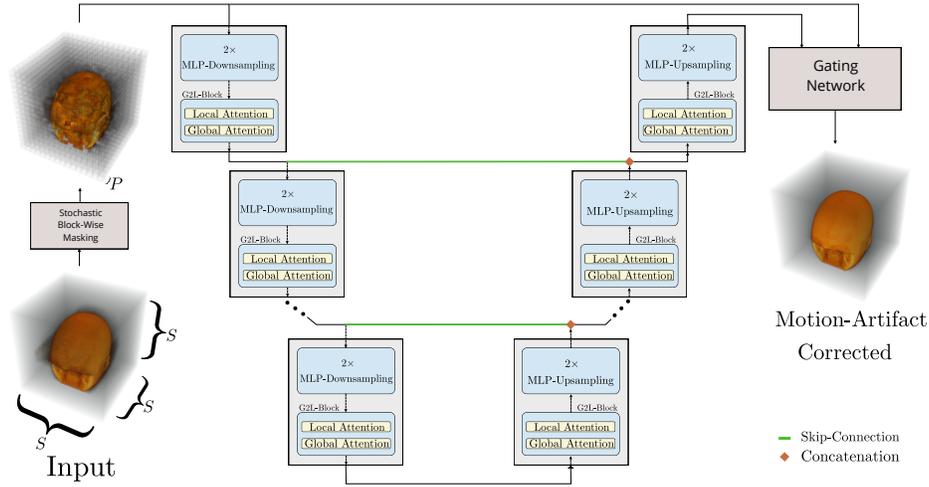}
\caption{\textbf{Architecture.} \model{MAMOC} arranges G2L transformer blocks in a U-Net-like structure, featuring a contracting encoder path and an expanding decoder path connected by skip connections and a bottleneck layer.
During training and inference, inputs are randomly masked for reconstruction. 
A gating network learns which voxels from the masked input can be preserved and passed to the output.
}
\label{PAPER-D:fig:Architecture}
\end{figure}

\section{Method}

In the following section, we describe how we design \model{MAMOC} to effectively filter out motion artifacts, taking advantage of self-supervision and transfer learning on large, artifact-free datasets  to be able to learn to correct real motion artifacts from limited supervised data in \dataset{MR-ART}. 

\subsubsection*{Architecture}
We adopt a U-Net-like architecture \citep{ronneberger2015u}, featuring a contracting encoder path successively producing coarser resolutions, and an expanding decoder path connected by skip connections and a bottleneck layer (Fig.\ \ref{PAPER-D:fig:Architecture}).
Instead of using convolutional elements in our network, we draw inspiration from \model{Swin UNETR} \citep{hatamizadeh2022swin}, which uses Transformer blocks \citep{vaswani_attention_is_2017} in the encoder path but convolutional layers in the decoder path. We opt for a symmetric architecture where both encoder and decoder use \emph{global-to-local} (\model{G2L}) windowed-attention blocks \citep{vandergoten2022wide} to avoid dependence on an extensive layering for achieving substantial spatial context size. 
This approach not only effectively addresses the non-local nature of motion artifacts but also reduces memory requirements \citep{vandergoten2022wide}, as fewer blocks are necessary.
\model{G2L} blocks are resolution-preserving, so to downsample from a resolution $S^3$ to $(\nicefrac{S}{2})^3$, we subdivide the representation into non-overlapping groups of voxels of size $2^3$, and project the eight voxels to a single number via a (shared) fully-connected layer.

\subsubsection*{Self-Supervised Training}
We first pre-train \model{MAMOC} in a self-supervised manner using \emph{masked autoencoding} in a similar manner to  approaches like \model{MAE} \citep{he_masked_2021} and \model{Paella} \citep{rampas_novel_2023}.
We randomly mask portions of the input data by subdividing an input scan $x \in \mathbb{R}^{S \times S \times S}$ into non-overlapping blocks of size $B$.
We then sample a \emph{keep probability} $p$ for each scan in the batch from a pre-defined range of values (\ie the unit interval) and mask out \emph{exactly} $\ceil{(1-p) \cdot (\nicefrac{S}{B})^3 }$ blocks. 
To guarantee an exact number of masked-out blocks, rather than only in expectation, we generate random permutations over the set of block indices, followed by thresholding.
All voxels in a masked-out block are set to zero. 
The objective of the network is to reconstruct these hidden segments, encouraging the model to learn the underlying structure of clean MRI data. 
We calculate the reconstruction loss using $\mathcal{L}_2$.
This self-supervision technique facilitates the learning of rich data representations by challenging the model to predict missing information \citep{he_masked_2021}. 
We apply this process on three high-quality datasets that lack motion artifacts: \dataset{ADNI} \citep{Wyman2013}, \dataset{OASIS-3} \citep{LaMontagne}, and \dataset{IXI} \citep{ixi_dataset_2023}.

\subsubsection*{Transfer Learning}
While self-supervision enables the network to learn rich features of high-quality MRI scans, it does not provide the model with understanding of motion artifacts or how to remove them.
Thus, we fine-tune the model using real motion artifacts from \dataset{MR-ART}. 
We sample moderately and heavily motion-affected scans and train the model to reconstruct the clean (motion-free) scan of the same subject. 
We continue to employ masking in this phase for two reasons: \emph{(1)} it acts as a regularization, improving performance and stabilizing training; \emph{(2)} masked reconstruction allows the model to learn to handle small misalignments in registration between the clean scan and motion-affected scans to improve the restoration quality.

\subsubsection*{Test-Time Prediction}
We propose a type of test-time augmentation to improve reconstruction in \model{MAMOC}.
At inference, rather than fixing the keep probability to $1.0$, we choose a specific $p$ (constrained to the interval defined before) and perform $L$ passes per scan.
We then average the predictions. 
For practical purposes, this is equivalent to repeating a single scan $L$ times over the batch dimension, so in effect the the throughput for a single scan is unchanged if sufficient memory is available. 
By default, we set $p=0.6$.
Empirical tests to determine this hyperparameter are provided in the \emph{Appendix}. 

%
%
%

\begin{figure}[t]
    \centering
    
    \begin{subfigure}[t]{\textwidth}
        \centering
        \caption*{\textbf{(a)}}
        \resizebox{\textwidth}{!}{
        \begin{tabular}[t]{lcccc}
        \toprule
        \multicolumn{1}{c}{Method} & \multicolumn{2}{c}{$ 128^3 $} & \multicolumn{2}{c}{$ 256^3 $} \\
        \cmidrule(l{3pt}r{3pt}){1-1} \cmidrule(l{3pt}r{3pt}){2-3} \cmidrule(l{3pt}r{3pt}){4-5}
        \multicolumn{1}{c}{} & \multicolumn{1}{c}{PSNR $\uparrow$} & \multicolumn{1}{c}{SSIM $\uparrow$} & \multicolumn{1}{c}{PSNR $\uparrow$} & \multicolumn{1}{c}{SSIM $\uparrow$} \\
        \cmidrule(l{3pt}r{3pt}){2-2} \cmidrule(l{3pt}r{3pt}){3-3} \cmidrule(l{3pt}r{3pt}){4-4} \cmidrule(l{3pt}r{3pt}){5-5}
        \model{AntsPy} & $ 24.841 \pm 1.407 $ (***) & $ 0.863 \pm 0.025 $ (***) & $ 27.350 \pm 1.652 $ (***) & $ 0.880 \pm 0.029 $ (***) \\
        \model{MCFLIRT} & $ 32.492 \pm 1.332 $ (***) & $ 0.942 \pm 0.028 $ (***) & $ 35.374 \pm 2.407 $ (*) & $ 0.945 \pm 0.033 $ (***) \\
        \model{V-Net} & $ 34.680 \pm 0.924 $ (***) & $ 0.845 \pm 0.063 $ (***) & $ 34.431 \pm 1.675 $ (***) & $ 0.602 \pm 0.063 $ (***) \\
        \model{W-G2L-ART} & $ 36.082 \pm 1.946 $ (***) & $ 0.979 \pm 0.010 $ (ns) & $ 36.034 \pm 1.956 $ (ns) & $ \mathbf{0.963 \pm 0.014} $\\
        \model{MAMOC} & $ \mathbf{37.118 \pm 1.683} $ & $ \mathbf{0.980 \pm 0.011} $ & $ \mathbf{36.041 \pm 1.977} $ & $ 0.959 \pm 0.020 $ (ns)\\
        \bottomrule
        \end{tabular}
        }
    \end{subfigure}
    
    \vspace{0.2cm} 
    
    \begin{subfigure}[t]{\textwidth}
        \centering
        \caption*{\textbf{(b)}}
        \resizebox{\textwidth}{!}{
        \begin{tabular}[t]{lcccc}
        \toprule
        \multicolumn{1}{c}{Method} & \multicolumn{2}{c}{$ 128^3 $} & \multicolumn{2}{c}{$ 256^3 $} \\
        \cmidrule(l{3pt}r{3pt}){1-1} \cmidrule(l{3pt}r{3pt}){2-3} \cmidrule(l{3pt}r{3pt}){4-5}
        \multicolumn{1}{c}{} & \multicolumn{1}{c}{SNR $\uparrow$} & \multicolumn{1}{c}{CNR $\uparrow$} & \multicolumn{1}{c}{SNR $\uparrow$} & \multicolumn{1}{c}{CNR $\uparrow$} \\
        \cmidrule(l{3pt}r{3pt}){2-2} \cmidrule(l{3pt}r{3pt}){3-3} \cmidrule(l{3pt}r{3pt}){4-4} \cmidrule(l{3pt}r{3pt}){5-5}
        \model{AntsPy} & $ 6.785 \pm 0.563 $ (***) & $ 1.452 \pm 0.175 $ (***) & $ 6.950 \pm 0.378 $ (***) & $ 1.675 \pm 0.210 $ (***)\\
        \model{MCFLIRT} & $ \mathbf{7.297 \pm 0.453} $ & $ 1.550 \pm 0.149 $ (***) & $ 7.297 \pm 0.501 $ (***) & $ 1.787 \pm 0.221 $ (**)\\
        \model{V-Net} & $ 6.924 \pm 0.302 $ (***) & $ 1.374 \pm 0.100 $ (***) & $ 6.310 \pm 0.364 $ (***) & $ 1.565 \pm 0.128 $ (***)\\
        \model{W-G2L-ART} & $ 6.946 \pm 0.471 $ (***) & $ 1.473 \pm 0.147 $ (***) & $ 6.974 \pm 0.405 $ (***) & $ 1.585 \pm 0.165 $ (***)\\
        \model{MAMOC} & $ 7.222 \pm 0.494 $ (ns) & $ \mathbf{1.648 \pm 0.120} $ & $ \mathbf{7.682 \pm 0.377} $ & $ \mathbf{1.800 \pm 0.147} $\\
        \bottomrule
        \end{tabular}
        }
    \end{subfigure}
    
    \caption{\textbf{(a)}  We compare how well motion-corrected scans produced by each method are able to reconstruct motion-free scans on \dataset{MR-ART}. \textbf{(b)} Signal-to-noise ratios (SNR) and contrast-to-noise ratios (CNR) as estimated by MRIQC (Markers for statistical significance: '***' for $p < 0.01$, '**' for $p < 0.05$, '*' for $p < 0.1$, 'ns' for $p \geq 0.1$) }
    \label{fig:paperD_combined_tables}
\end{figure}

\section{Experiments}
 
\subsubsection*{Baseline Methods}
We benchmark \model{MAMOC} against both traditional and deep learning-based artifact correction methods.
\model{MCFLIRT}  \citep{jenkinson2002improved} is a widely-used MC tool 
that accepts both fMRI and structural MRI scans. \model{ANTsPy}  \citep{tustison2021antsx} is a recently proposed MRI framework that offers MC capabilities. 
\model{W-G2L-ART}  \citep{vandergoten2022wide} is a Transformer-based artifact correction architecture that uses a wide range of synthetic artifact classes to enable supervised learning. 
Furthermore, in order to rule out that the MC task can effectively be addressed by merely training in a supervised manner on \dataset{MR-ART}, we compare \model{MAMOC} against a \model{V-Net}  \citep{milletari_v-net}.
The \model{V-Net} is trained by using motion-affected \dataset{MR-ART} scans as inputs and is tasked to minimize the $\mathcal{L}_2$ error \wrt to the clean, same-subject \dataset{MR-ART} scan. 
\model{W-G2L-ART} is trained using synthetic artifacts on \dataset{ADNI}, \dataset{OASIS-3}, and \dataset{IXI}.
\model{MCFLIRT} and \model{ANTsPy} use rigid and non-rigid body transformations and as such do not require training data.

%
%
%

\subsubsection*{Data \& Preprocessing}
We standardized the resolution of all datasets to $256^3$, equivalent to a voxel size of $1\text{mm}^3$, to ensure consistency across the board.
Because \dataset{MR-ART} has been de-identified using a defacing technique (PyDeface  \citep{omer-faruk-gulban-2022}), we apply the very same method to \dataset{ADNI}, \dataset{OASIS-3} and \dataset{IXI} for the sake of consistency.
The \dataset{ADNI} \citep{Wyman2013} data used in the preparation of this article were obtained from the Alzheimer’s Disease
Neuroimaging Initiative database (\url{adni.loni.usc.edu}). \dataset{ADNI} was launched in
2003 as a public-private partnership, led by Principal Investigator Michael W. Weiner,
MD. The primary goal of \dataset{ADNI} has been to test whether serial magnetic resonance imaging
(MRI), positron emission tomography (PET), other biological markers, and clinical and
neuropsychological assessment can be combined to measure the progression of mild
cognitive impairment (MCI) and early Alzheimer’s disease (AD). For up-to-date information,
see \url{www.adni-info.org}.
A complication with the \dataset{MR-ART} dataset is the lack of spatial alignment among scans of the same subject, complicating the application of traditional voxel-to-voxel loss metrics.
To mitigate this issue, we adopted a tailored registration approach: a two-stage process for \dataset{ADNI}, \dataset{OASIS-3} and \dataset{IXI}, and a more complex three-stage process for \dataset{MR-ART}.
This involved initially co-registering any three scans from the same subject within \dataset{MR-ART}, followed by intra-dataset registration across different subjects. 
The final phase involved co-registering scans across all datasets to achieve uniformity. We use the SyN algorithm  \citep{avants2008symmetric}, a non-linear diffeomorphic registration method in all steps.
In order to do away with the need of having to rely on a single fixed template, we use the template-building algorithm from ANTs  \citep{avants2010optimal} 
on the to-be-registered scans prior to co-registration. All scans are subsequently intensity normalized across all datasets using least squares normalization  \citep{reinhold2019evaluating}.
In a final step, we again resample the scans to resolution $128^3$ (voxel size $2\text{mm}^3$) to allow additional comparisons. 
We randomly partitioned \dataset{MR-ART} into a $70$-$30$ train-test split by subjects.
 
To enable a fair comparison, we scale the model complexity of each model to fully occupy the 24 GiB GPU memory of an NVIDIA RTX 4090 with a batch size of $4$. 
To reduce memory requirements, we apply the following techniques: Training with automatic mixed precision \citep{mici_mixed_2018}, graph compilation via the \emph{Inductor} backend \citep{ansel_torch_2024}, and using the memory-friendly Lion  \citep{chen2023symbolic} optimizer.

\begin{figure}[t]
\centering
\begin{minipage}{0.13\textwidth}
    \centering
    \text{\small{Input}}
\end{minipage}\hfill
\begin{minipage}{0.13\textwidth}
    \centering
    \text{\small{AntsPy}}
\end{minipage}\hfill
\begin{minipage}{0.13\textwidth}
    \centering
    \text{\small{MCFLIRT}}
\end{minipage}\hfill
\begin{minipage}{0.13\textwidth}
    \centering
    \text{\small{V-Net}}
\end{minipage}\hfill
\begin{minipage}{0.13\textwidth}
    \centering
    \text{\scriptsize{W-G2L-ART}}
\end{minipage}\hfill
\begin{minipage}{0.13\textwidth}
    \centering
    \text{\small{MAMOC}}
\end{minipage}\hfill
\begin{minipage}{0.13\textwidth}
    \centering
    \text{\small{GT}}
\end{minipage}\\
\includegraphics[width=0.13\textwidth]{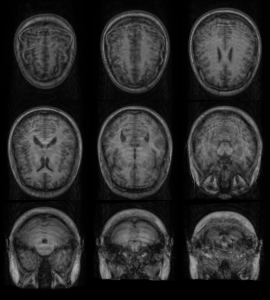}\hfill
\includegraphics[width=0.13\textwidth]{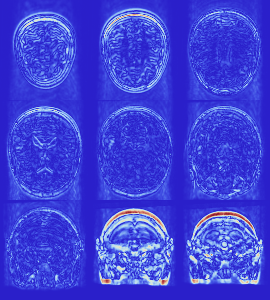}\hfill
\includegraphics[width=0.13\textwidth]{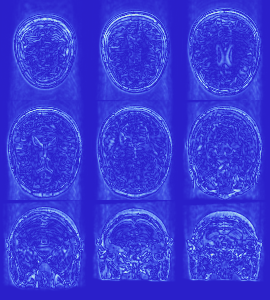}\hfill
\includegraphics[width=0.13\textwidth]{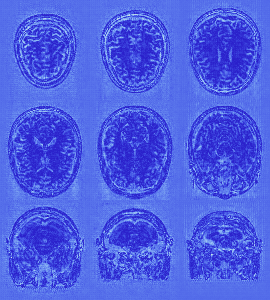}\hfill
\includegraphics[width=0.13\textwidth]{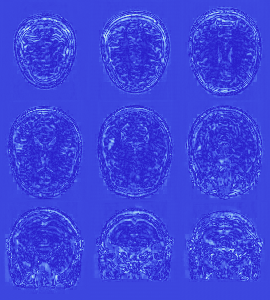}\hfill
\includegraphics[width=0.13\textwidth]{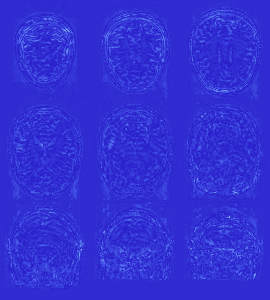}\hfill
\includegraphics[width=0.13\textwidth]{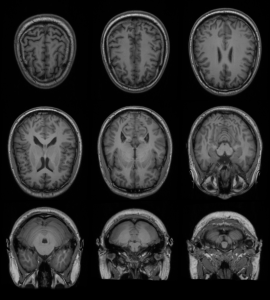}\hfill\\
\includegraphics[width=0.13\textwidth]{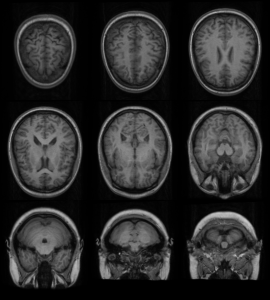}\hfill
\includegraphics[width=0.13\textwidth]{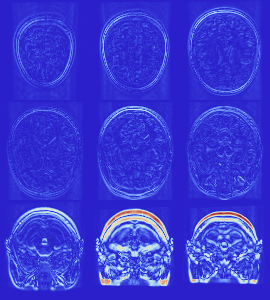}\hfill
\includegraphics[width=0.13\textwidth]{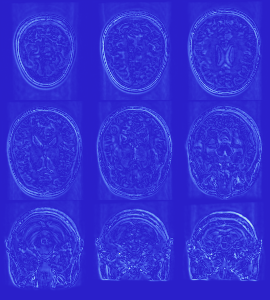}\hfill
\includegraphics[width=0.13\textwidth]{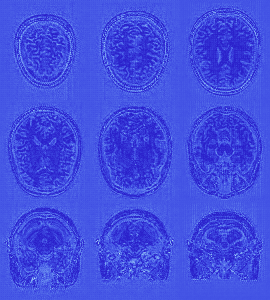}\hfill
\includegraphics[width=0.13\textwidth]{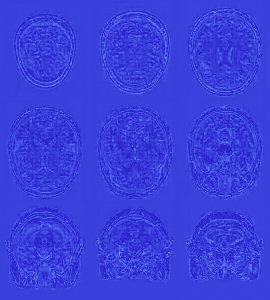}\hfill
\includegraphics[width=0.13\textwidth]{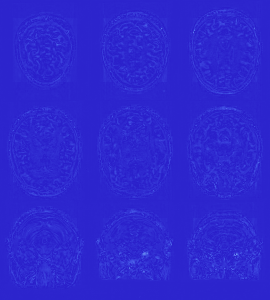}\hfill
\includegraphics[width=0.13\textwidth]{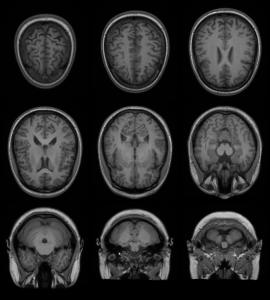}\hfill\\
\includegraphics[width=0.13\textwidth]{figs/diff_maps/1-2.png}\hfill
\includegraphics[width=0.13\textwidth]{figs/diff_maps/2-2.png}\hfill
\includegraphics[width=0.13\textwidth]{figs/diff_maps/3-2.png}\hfill
\includegraphics[width=0.13\textwidth]{figs/diff_maps/4-2.png}\hfill
\includegraphics[width=0.13\textwidth]{figs/diff_maps/5-2.png}\hfill
\includegraphics[width=0.13\textwidth]{figs/diff_maps/6-2.png}\hfill
\includegraphics[width=0.13\textwidth]{figs/diff_maps/7-2.png}\hfill\\
\includegraphics[width=0.3\textwidth]{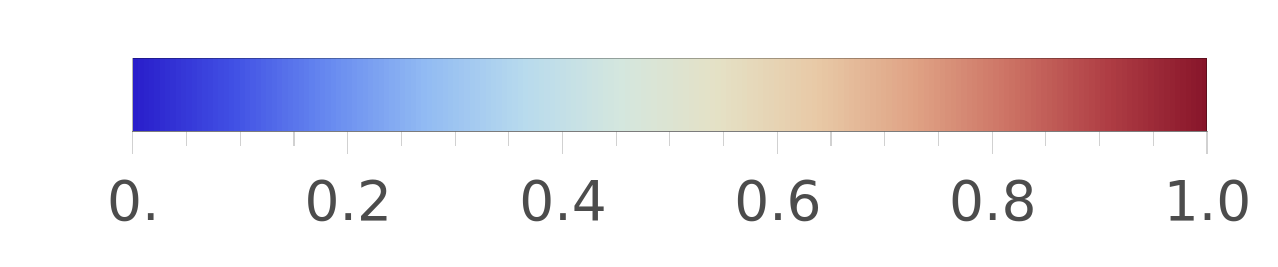}
\caption{\textbf{Difference Maps.} We visualize the reconstruction quality with absolute difference between the outputs of various models and the ground-truth (GT), \ie motion-free scans, given the same motion-affected inputs. Each row of difference maps is normalized as a group \wrt the unit interval, with lower values indicating better reconstruction quality and greater similarity to the ground truth.}
\label{PAPER-D:fig:diffMaps}
\end{figure}

\begin{figure}[t]
    \centering
    \includegraphics[width=0.75\textwidth]{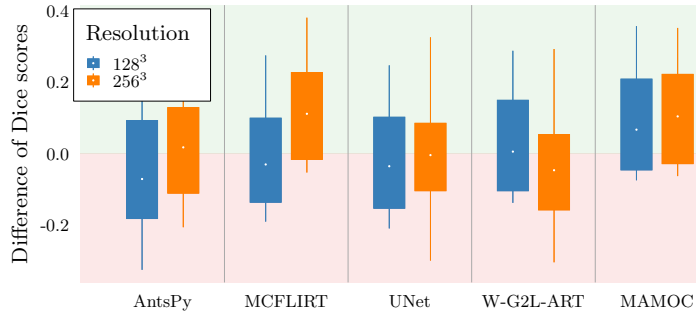}
    \caption{\textbf{Subcortical Brain Segmentation.} Positive values indicate that segmentation with FASTSURFER benefits from applying an MC algorithm on motion-affected scans. The $y$-axis denotes to which extent the class-averaged (96 classes) Dice score improves if an motion-affected scan is motion corrected. See text for details.}
    \label{PAPER-D:fig:fastsurfer}
\end{figure}

\subsubsection*{Reconstruction Quality Analysis}
To measure reconstruction quality we compare the motion-corrected output from the various models to the motion-free scan from the \dataset{MR-ART} test set.
We report the \emph{peak signal-to-noise ratio} (PSNR) as well as the \emph{structural similarity index} (SSIM), two metrics commonly used to quantify visual reconstruction quality. 
The results reported in Fig.~\ref{fig:paperD_combined_tables} \textbf{(a)} showcases that \model{MAMOC} outperforms the other methods, in most cases by a significant margin.
The lone exception is W-G2L-ART on $256^3$ resolution volumes, as measured by PSNR, which outperforms \model{MAMOC} by a very small margin.
Visual depictions of the reconstruction quality as well as difference maps between the output of the various correction methods and the motion-free scan are provided in Fig. \ref{PAPER-D:fig:diffMaps}.

\subsubsection*{Image Quality Analysis}
We use the \emph{MRI Quality Control} (MRIQC) framework  \citep{esteban2017mriqc} to gauge the image quality of motion-affected, motion-free and motion-corrected scans. 
The framework automatically generates a wide array of image-quality metrics specifically for MRI scans.
We report the \emph{signal-to-noise} ratio (SNR) and the \emph{contrast-to-noise} (CNR) ratio as estimated by MRIQC in Fig.~\ref{fig:paperD_combined_tables} \textbf{(b)}.
The latter ratio measures the separation of intensity distributions corresponding to gray and white matter. 
We observe that \model{MAMOC} attains the highest SNR value with respect to the $256^3$ resolution and is only second to \model{MCFLIRT} in terms of the $128^3$ resolution.

\subsubsection*{Subcortical Brain Segmentation}
To further validate the efficacy of \model{MAMOC}, we investigate its impact on the downstream task of brain segmentation. 
We utilize FASTSURFER  \citep{HENSCHEL2020117012}, a state-of-the-art deep learning approach for subcortical brain segmentation, and apply it to three distinct MRI scans of the same subject: a motion-free scan ($x_\text{MF}$), a motion-affected scan ($x_\text{MA}$), and a motion-corrected scan ($x_\text{MC}$).
The metric $\text{Dice}(x_\text{MF}, x_\text{MC}) - \text{Dice}(x_\text{MF}, x_\text{MA})$ is employed to quantify the enhancement in segmentation performance when using the motion-corrected scan compared to the motion-affected scan. 
The results are presented in Figure \ref{PAPER-D:fig:fastsurfer}. While some motion correction methods fail to consistently improve downstream segmentation performance, \model{MAMOC} is notably the only method that consistently improves improves brain segmentation outcomes in the presence of motion artifacts. This is important since subcortical brain segmentation deals with the identification of small structures that can easily be missed if a scan is affected by artifacts.

\section{Conclusion}

Our study introduced \model{MAMOC}, a self-supervised approach employing masked autoencoding for MRI motion artifact correction.
This work stands out as the first to assess motion correction techniques against real-world motion artifacts on the \model{MR-ART} dataset. 
The results demonstrate \model{MAMOC}'s efficacy in generating high-resolution, motion-corrected MRI images and its superior performance over existing methods across various metrics, including PSNR and SSIM, and furthermore in enhancing subcortical brain segmentation.
A key aspect of \model{MAMOC} is its use of masked autoencoding not only during training but also at inference time, which contributes to improved reconstruction quality. 
This approach, combined with the ability to process images at native resolution, ensures high-quality correction of motion artifacts.
\model{MAMOC} can, in principle, also be trained for use in other volumetric medical imaging modalities such as CT/EEG or PET.
Its efficiency on consumer-grade hardware and applicability to various imaging modalities underscore its potential for broader clinical and research applications.

\subsection*{Acknowledgements}
\footnotesize
\textit{
Data collection and sharing for this project was funded by the Alzheimer's Disease
Neuroimaging Initiative (ADNI) (National Institutes of Health Grant U01 AG024904) and
DOD ADNI (Department of Defense award number W81XWH-12-2-0012). ADNI is funded
by the National Institute on Aging, the National Institute of Biomedical Imaging and
Bioengineering, and through generous contributions from the following: AbbVie, Alzheimer’s
Association; Alzheimer’s Drug Discovery Foundation; Araclon Biotech; BioClinica, Inc.;
Biogen; Bristol-Myers Squibb Company; CereSpir, Inc.; Cogstate; Eisai Inc.; Elan
Pharmaceuticals, Inc.; Eli Lilly and Company; EuroImmun; F. Hoffmann-La Roche Ltd and
its affiliated company Genentech, Inc.; Fujirebio; GE Healthcare; IXICO Ltd.; Janssen
Alzheimer Immunotherapy Research \& Development, LLC.; Johnson \& Johnson
Pharmaceutical Research \& Development LLC.; Lumosity; Lundbeck; Merck \& Co., Inc.;
Meso Scale Diagnostics, LLC.; NeuroRx Research; Neurotrack Technologies; Novartis
Pharmaceuticals Corporation; Pfizer Inc.; Piramal Imaging; Servier; Takeda Pharmaceutical
Company; and Transition Therapeutics. The Canadian Institutes of Health Research is
providing funds to support ADNI clinical sites in Canada. Private sector contributions are
facilitated by the Foundation for the National Institutes of Health (www.fnih.org). The grantee
organization is the Northern California Institute for Research and Education, and the study is
coordinated by the Alzheimer’s Therapeutic Research Institute at the University of Southern
California. ADNI data are disseminated by the Laboratory for Neuro Imaging at the
University of Southern California.}
\normalsize

\bibliographystyle{abbrvnat}
\bibliography{bibliography}


\end{document}